\documentclass[twocolumn,showpacs,amsmath,amssymb,aps,prd,nofootinbib]{revtex4}

\usepackage{graphicx}
\usepackage{color} 
\usepackage{soul}  

\begin{document}

\title{Gravitational-wave generation in hybrid quintessential inflationary models}

\author{Paulo M. S\'a}

\email{pmsa@ualg.pt}

\affiliation{Departamento de F\'{\i}sica, Faculdade de Ci\^encias
e Tecnologia, Universidade do Algarve, Campus de Gambelas,
8005-139 Faro, Portugal}

\author{Alfredo B. Henriques}

\email{alfredo.henriques@ist.utl.pt}

\affiliation{Centro Multidisciplinar de Astrof\'{\i}sica -- CENTRA and
Departamento de F\'{\i}sica, Instituto Superior T\'ecnico, UTL, Av.\ Rovisco
Pais, 1049-001 Lisboa, Portugal}

\date{March 18, 2010}

\begin{abstract}
We investigate the generation of gravitational waves in the hybrid
quintessential inflationary model. The full gravitational-wave energy spectrum
is calculated using the method of continuous Bogoliubov coefficients. The
post-inflationary kination period, characteristic of quintessential
inflationary models, leaves a clear signature on the spectrum, namely, a peak
at high frequencies. The maximum of the peak is firmly located at the MHz-GHz
region of the spectrum and corresponds to $\Omega_\textsc{gw} \simeq10^{-12}$.
This peak is substantially smaller than the one appearing in the
gravitational-wave energy spectrum of the original quintessential inflationary
model, therefore avoiding any conflict with the nucleosynthesis constraint on
$\Omega_\textsc{gw}$.
\end{abstract}


\maketitle

\section{Introduction}

Gravitational waves of cosmological origin are at present the
object of an important research effort, as people realize that
they will provide us with a unique telescope to the very early
Universe, giving us information not easily available by any other
means. Indeed, it would be of the utmost importance to be able to
find direct signals coming from these earliest of times, in order
to improve our understanding about inflation, preheating and
reheating mechanisms, post-inflationary phase transitions,
topological defects of grand unified theories, and string theory,
among other issues relevant for both cosmology and high-energy
physics.

In this work we investigate gravitational-wave generation within a
model recently proposed by Bastero-Gil \emph{et
al.}~\cite{bastero-gil-et-al}, a modification of the original
quintessential inflationary model of Peebles and Vilenkin
\cite{peebles-vilenkin}. In the original model, a unified
description of inflation and dark energy is achieved with a single
scalar field $\phi$ playing both roles of inflaton and
quintessence. In such quintessential inflationary models reheating
does not proceed in the usual way, through the complete decay of
the scalar field into quanta of other fields. In fact, since the
potential of the scalar field  has no minimum, inflation is
followed not by coherent oscillations of the scalar field, but
rather by a kination period \cite{spokoiny,joyce}, during which
the evolution is dominated by the kinetic energy of the scalar
field $\phi$, which behaves approximately as stiff matter.
Reheating proceeds then by gravitational particle production
taking place at the transition from the inflationary to the
kination period \cite{ford}. This mechanism is, however, quite
inefficient and may lead to cosmological problems associated with
large isocurvature fluctuations and overproduction of gravitinos
and moduli fields \cite{felder-kofman-linde}. In order to recover
the usual reheating mechanism, Bastero-Gil \emph{et
al.}~\cite{bastero-gil-et-al} introduce another scalar field
$\chi$, coupled to the original inflaton/quintessence field
$\phi$, with a hybrid-like potential \cite{linde}. In such a
hybrid quintessential inflationary model, during the kination
period driven by the kinetic energy of the field $\phi$, the field
$\chi$ oscillates around the minimum of the hybrid-like potential,
completely decaying into relativistic particles, thus reheating
the Universe. Since this reheating mechanism is more efficient
than gravitational particle production, the radiation-dominated
epoch of expansion of the Universe begins earlier and, therefore,
the kination period is shorter in this hybrid model than in the
original one.

It is known that a phase of evolution in the early Universe with
an equation of state stiffer than radiation leads to a sharp
increase of the gravitational-wave spectral energy density
parameter $\Omega_\textsc{gw}$ in the high-frequency region of the
spectrum \cite{grishchuk77, hu-parker, giovannini98}. This peak in
the gravitational-wave spectrum is unavoidable in quintessential
inflationary models, in which inflation is followed by a period of
kination \cite{giovannini99, riazuelo-uzan, sahni-sami-souradeep,
tashiro-chiba-sasaki}. In the original quintessential inflationary
model of Peebles and Vilenkin \cite{peebles-vilenkin}, the height
of the peak depends on the number of scalar degrees of freedom,
$N_s$, whose decay into fermions triggers the onset of a
gravitational reheating of the Universe. For the minimum allowed
value of $N_s$, of the order of $10^2$, the gravitational-wave
spectral energy density parameter $\Omega_\textsc{gw}$ could be as
high as $10^{-6}$ \cite{giovannini99}, several orders of magnitude
higher than in standard inflation. Clearly, in the hybrid
quintessential inflationary model, in which the kination period is
shorter than in the original model, one expects the high-frequency
peak of the spectrum to be smaller.

Our main aim in this paper will be to calculate, within the hybrid
quintessential inflationary model, the present-day spectrum of the
cosmological gravitational waves generated during the evolution of
the Universe and to show that its careful analysis will allow us
to extract much information about the properties of the potential
driving both the primordial inflationary stage and the present-day
accelerated expansion. To take into account the expansion of the
Universe during the transitions between the different stages of
its evolution, characterized by different equations of state, we
use throughout the method first developed by Parker, in his
seminal paper of 1969~\cite{parker}. This method, applied by
Parker to particle production in an expanding Universe and then
extended to the case of gravitons
\cite{henriques94,henriques-moorhouse-mendes,mendes-henriques-moorhouse},
is based on the time evolution of the Bogoliubov coefficients,
which obey appropriate differential equations. The numerical
integration of these equations will immediately allow us to
construct the full spectrum of the gravitational waves (see
Refs.~\cite{henriques04,sa-henriques1,sa-henriques2,sa-henriques-potting}
for applications of this method to several cosmological models).

Our paper is organized as follows. In the next section we describe
the hybrid quintessential inflationary model of Bastero-Gil
\emph{et al.}~\cite{bastero-gil-et-al} and write the equations of
motion for the different stages of the expansion. We also describe
the simple phenomenological mechanism responsible for the
reheating of the Universe, acting during the kination period that
follows inflation. In section~\ref{sect-gravitational-waves}, we
present the aforementioned method of the continuous Bogoliubov
coefficients to calculate the spectrum of the gravitational waves,
integrate the corresponding equations and compute the
gravitational-wave spectral energy density parameter for
frequencies ranging from about $10^{-17}\mbox{ rad/s}$ to about
$10^{10}\mbox{ rad/s}$. The influence on the spectrum of the
various parameters defining the hybrid quintessential inflationary
potential and the resulting constraints are then carefully
analyzed in the different frequency regions. In
section~\ref{sect-conclusions} we summarize our main conclusions.

\section{Hybrid Quintessential Inflation\label{sect-hybrid-quint-inflation}}

\subsection{The hybrid potential}

As shown by Peebles and Vilenkin \cite{peebles-vilenkin}, the unification of
inflation and dark energy within a single framework, which they termed
\textit{quintessential inflation}, can be achieved with a single scalar field
with potential given by
\begin{eqnarray}
V(\phi) = \left\{
\begin{tabular}{ll}
 $\lambda_\phi (\phi^4+M^4)$, & \quad \mbox{for} \quad $\phi <0$, \\
 \\
 $\lambda_\phi M^8(\phi^4+M^4)^{-1}$,  & \quad \mbox{for} \quad
 $\phi\geqslant0$,
\end{tabular}
\right. \label{potential-V}
\end{eqnarray}
where $\lambda_\phi$ and $M$ are constants. The value of
$\lambda_\phi$ is constrained by recent measurements of the cosmic
microwave background and large-scale structure to be of the order
of $10^{-13}$ (see Ref.~\cite{smith-kamionkowski-cooray} for a
derivation of such constraints for different models of inflation).
On the other hand, agreement with the measured value of today's
dark-energy density \cite{wmap7} requires $M$ to be of the order
of $10^{-14}\, m_{\mbox{\tiny p}}$.

\begin{figure}[t]
\includegraphics[width=8.1cm]{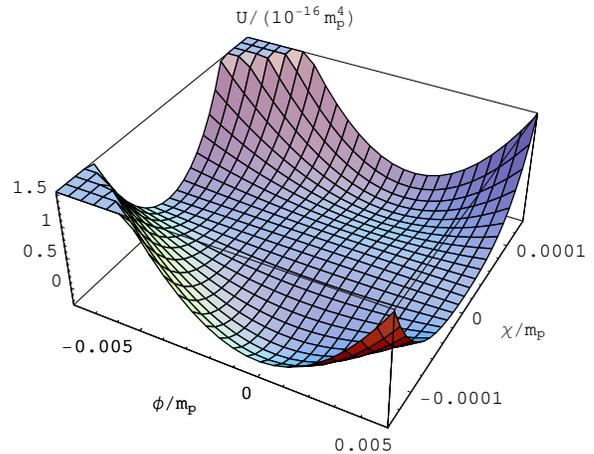}
\caption{Potential $U(\phi,\chi)$ for $m=0.004\, m_{\mbox{\tiny p}}$,
$\lambda_\phi = 10^{-13}$, $M = 1.7 \times 10^{-14}\, m_{\mbox{\tiny p}}$,
$\lambda_\chi=1$ and $g=0.025$.} \label{fig-potential3D}
\end{figure}

Recently, Bastero-Gil \emph{et al.}~\cite{bastero-gil-et-al}
proposed a modification of the original quintessential
inflationary model, in which the reheating of the Universe is
driven not by gra\-vi\-tational particle production but rather by
the simpler mechanism of the decay of an oscillating massive
field. This is achieved by the introduction of an extra scalar
field $\chi$ with a hybrid-like potential given by
\begin{eqnarray}
U(\phi,\chi) = V(\phi) + \frac12 g^2 \chi^2 (\phi^2-m^2)+\frac14 \lambda_\chi
\chi^4, \label{potential-U}
\end{eqnarray}
where $V(\phi)$ is given by Eq.~(\ref{potential-V}) and $g$, $m$, and
$\lambda_\chi$ are some parameters to be specified. Note that for
$|\phi|\geqslant m$ the potential (\ref{potential-U}) has just one minimum at
$\chi=0$, while for $|\phi|< m$ it has two minima located at
$\chi=\pm\sqrt{g^2(m^2-\phi^2)/\lambda_\chi}$ (see
Fig.~\ref{fig-potential3D}).

In our analysis of this hybrid quintessential inflationary model we will
divide the evolution of the Universe into four stages.

In the first stage of evolution, $\phi\leqslant-m$ and $\chi$ is
assumed to be at rest at the minimum of the potential ($\chi=0$).
Initial conditions are such that the potential energy of the
scalar field $\phi$ dominates the evolution of the Universe,
yielding a period of chaotic inflation. By the end of this stage
of evolution, the kinetic term in the energy density of $\phi$
becomes dominant, giving rise to a period of kination, during
which the scalar field behaves as stiff matter with equation of
state $p=w\rho$, $w=1$.

During the second stage of evolution, which takes place for
$|\phi|<m$, the origin of the potential becomes unstable for the
scalar field $\chi$. As a result, this field rolls down toward one
of the new minima of the potential, located at
$\chi=\pm\sqrt{g^2(m^2-\phi^2)/\lambda_\chi}$, and starts to
oscillate around it. Of course, this behavior of the scalar field
$\chi$ is expected only if the duration of the second stage of
evolution is much greater than the characteristic response time of
the field to changes in the potential. As shown in
Ref.~\cite{bastero-gil-et-al}, this requirement can be formulated
as a constraint on the values of the parameters $g$ and $m$,
namely,
\begin{eqnarray}
g \left( \frac{m}{m_{\mbox{\tiny p}}} \right)^2 \gtrsim 0.2 \,
\lambda_\phi^{1/2}. \label{constraint-1}
\end{eqnarray}
We also demand that the motion of the scalar field $\phi$ is not influenced
significantly by $\chi$, i.e., we impose the condition that the energy density
of $\chi$ is much smaller than the kinetic term in the energy density of
$\phi$ during the second stage of evolution. This condition can be also
translated into a constraint on the values of the parameters appearing in the
potential~(\ref{potential-U}), namely \cite{bastero-gil-et-al},
\begin{eqnarray}
\frac{g^4}{\lambda_\chi} \left( \frac{m}{m_{\mbox{\tiny p}}} \right)^4
\lesssim 0.001 \, \lambda_\phi. \label{constraint-2}
\end{eqnarray}
Note that this last condition implies that kination extends throughout the
second stage of evolution.

The third stage of evolution takes place for $\phi\geqslant m$.
For such values of the scalar field $\phi$, the potential
(\ref{potential-U}) has again only one minimum in the $\chi$
direction, located at the origin. The scalar field $\chi$
oscillates around this stable minimum, transferring its kinetic
energy, acquired during the second stage of evolution, to a
radiation fluid, thus reheating the Universe. This decay of the
oscillating scalar field $\chi$ into radiation is achieved by the
introduction of a phenomenological dissipative coupling, between
the scalar field $\chi$ and the radiation fluid, proportional to
the mass of $\chi$ \cite{bastero-gil-et-al,yokoyama-maeda},
\begin{eqnarray}
\Gamma_\chi = \mu m_\chi = \mu g \sqrt{\phi^2-m^2}, \label{Gamma_chi}
\end{eqnarray}
$\mu$ being the proportionality constant. The above-mentioned
condition that the motion of the scalar field $\phi$ is not
influenced significantly by $\chi$ also imposes a constraint on
the values of $\mu$, namely \cite{bastero-gil-et-al},
\begin{eqnarray}
\mu \gg  0.2 \, \lambda_\phi^{1/6} \left( \frac{m_{\mbox{\tiny p}}}{m}
\right)^{2/3} \frac{g^{5/3}}{\lambda_\chi}. \label{constraint-3}
\end{eqnarray}
Note that this constraint, as well as constraints
(\ref{constraint-1}) and (\ref{constraint-2}), was derived
assuming that $m\ll m_{\mbox{\tiny p}}$\footnote{The constraints
(\ref{constraint-1}), (\ref{constraint-2}), and
(\ref{constraint-3}) differ from the ones derived in
Ref.~\cite{bastero-gil-et-al} by some numerical factors. This is
due to the fact that Bastero-Gil \emph{et al.}~estimated the
scale-factor growth between the end of the inflationary period and
the beginning of the second stage of evolution to be of the order
of 8, while our numerical simulations show this growth to be half
of this value. We take into account this factor 2, since the
constraints involve third and sixth powers of the scale-factor
growth.}. During the third stage of evolution, the energy density
of $\chi$ decays away rapidly. Gradually, as the energy density of
radiation increases, kination gives place to a radiation-dominated
Universe.

The fourth stage of evolution extends from the beginning of the
radiation-dominated era to the present epoch. The scalar field
$\chi$ is assumed to have decayed away completely in the previous
stage of evolution. A matter component (dark and usual, baryonic,
matter) is introduced into the equations of motion, giving rise to
a intermediate matter-dominated period in the evolution of the
Universe. The scalar field $\phi$, which in the first stage of
evolution played the role of inflaton, stays now practically
constant and, at late times, begins to dominate the evolution of
the Universe, giving rise to the present epoch of accelerated
expansion. Therefore, the scalar field $\phi$ plays, within this
model, both roles of inflaton and quintessence.

\subsection{The evolution of the Universe}

Let us now present the equations of motion for the different
stages of evolution.

For the first and second stages of evolution, the equations of
motion are
\begin{eqnarray}
\frac{\ddot{a}}{a} = -\frac{8\pi}{3m_{\mbox{\tiny p}}^2} \left( \dot{\phi}^2 +
\dot{\chi}^2 - U \right),  \label{adotdot12}
\end{eqnarray}
\vspace{-5mm}
\begin{eqnarray}
\ddot{\phi} + 3 \frac{\dot{a}}{a} \dot{\phi} + \frac{\partial U}{\partial
\phi}=0,
\end{eqnarray}
\vspace{-6mm}
\begin{eqnarray} \ddot{\chi} + 3 \frac{\dot{a}}{a} \dot{\chi} +
\frac{\partial U}{\partial \chi}=0,  \label{chidot12}
\end{eqnarray}
\vspace{-6mm}
\begin{eqnarray} \left( \frac{\dot{a}}{a} \right)^2 =
\frac{8\pi}{3m_{\mbox{\tiny p}}^2} \left( \frac{\dot{\phi}^2}{2} +
\frac{\dot{\chi}^2}{2} + U \right), \label{friedmann12}
\end{eqnarray}
where we have assumed a flat Friedmann-Robertson-Walker metric,
$a$ is the scale factor, $m_{\mbox{\tiny p}}$ is the Planck
mass\footnote{In the first, second, and third stages of evolution
we use the na\-tu\-ral system of units, with $\hbar=c=1$ and
$m_{\mbox{\tiny p}}=G^{-1/2}=1.22\times10^{19}\mbox{ GeV}$, while
in the fourth stage we use the International Systems of Units.},
and a dot denotes a derivative with respect to the cosmic time
$t$.

A Runge-Kutta method is used to solve the system of differential equations
(\ref{adotdot12})--(\ref{chidot12}), while Eq.~(\ref{friedmann12}) is used as
a constraint equation to check the accuracy of the numerical solution.

As initial condition for the scalar field $\phi$ we choose $\phi_{i1}=-5
\,m_{\mbox{\tiny p}}$, which guarantees enough inflation. The scalar field
$\chi$ is located near the origin; for numerical convenience, we assume
$\chi_{i1}=\dot{\chi}_{i1}=0$.  The initial values of the other variables are
chosen to be $a_{i1}=1$ and $\dot{\phi}_{i1}=\sqrt{2U(\phi_{i1},\chi_{i1})}$,
while $\dot{a}_{i1}$ is fixed by the Friedmann equation~(\ref{friedmann12}).
As mentioned above, the parameters $\lambda_\phi$ and $M$ of the potential
$V(\phi)$ are constrained by observations to be of the order of $10^{-13}$ and
$10^{-14}\, m_{\mbox{\tiny p}}$, respectively. The first stage of evolution
ends when $\phi=-m$; for the parameter $m$ we choose $m\lesssim 10^{-2} \,
m_{\mbox{\tiny p}}$.

\begin{figure}[t]
\includegraphics[width=8.1cm]{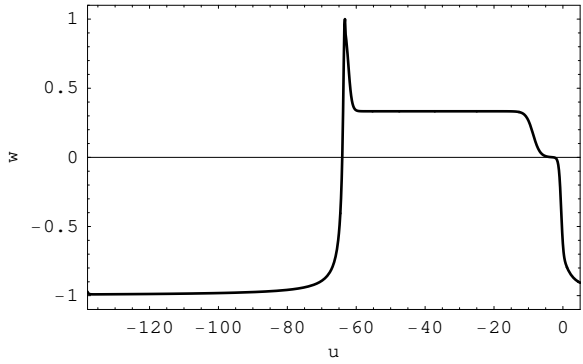}
\caption{Time evolution of the equation-of-state parameter
$w=p/\rho$, from the beginning of the inflationary period until
the present epoch, for $\lambda_\phi=10^{-13}$,
$M=1.7\times10^{-14}\, m_{\mbox{\tiny p}}$, $g=0.025$, $m=0.004\,
m_{\mbox{\tiny p}}$, $\lambda_\chi=1$, and $\mu=0.1$. In this
figure, instead of the cosmic time $t$, we use the variable
$u=-\ln(1+z)$, where $z$ is the redshift. The first stage of
evolution corresponds to $-139 \lesssim u \lesssim -63$, the
second stage to $u\approx -63$, the third stage to $-63 \lesssim u
\lesssim -60$, and the fourth stage to $-60\lesssim u \leqslant
0$. The epochs of inflation, kination, radiation domination,
matter domination, and dark-energy domination can be clearly
identified in this figure.} \label{fig-w-all-stages}
\end{figure}

For such values of the initial conditions and of the parameters
$\lambda_\phi$, $M$, and $m$, the first stage of evolution lasts for about
$10^7 \, t_{\mbox{\tiny p}}$ and the scale factor grows about $32$ orders of
magnitude. The equation-of-state parameter $w=p/\rho$ changes gradually from
$w=-1$ (inflation) to $w=1$ (kination), as shown in
Fig.~\ref{fig-w-all-stages}.

\begin{figure}[t]
\includegraphics[width=8.1cm]{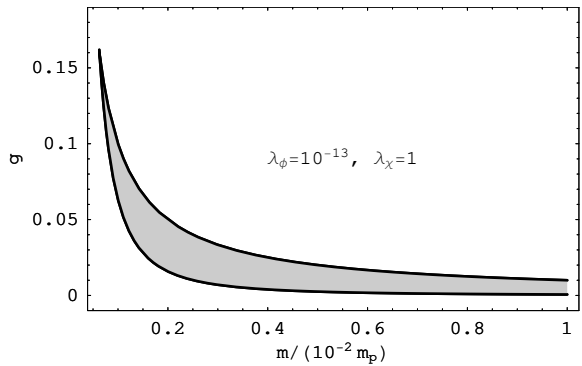}
\caption{Allowed values of the parameters $m$ and $g$ (shaded region), for
$\lambda_\phi=10^{-13}$ and $\lambda_\chi=1$.} \label{fig-constants}
\end{figure}

\begin{figure}[t]
\includegraphics[width=8.1cm]{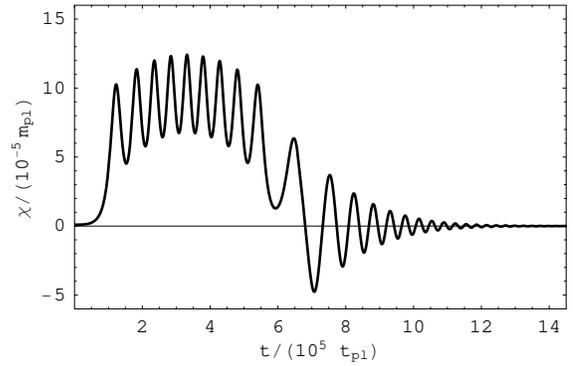}
\caption{Time evolution of the scalar field $\chi$ during the second and third
stages of evolution, for the same values of the parameters as in
Fig.~\ref{fig-w-all-stages}. The second stage of evolution lasts about
$7\times10^5 \, t_{\mbox{\tiny p}}$; the third stage of evolution, which lasts
about $5\times 10^{10} \, t_{\mbox{\tiny p}}$, is only partially shown is this
figure.} \label{fig-chi-23-etapas}
\end{figure}

At the beginning of the second stage of evolution, the origin becomes unstable
for the scalar field $\chi$. We perturb slightly this field, making it roll
towards the temporary minimum located at $\chi_{\mbox{\scriptsize min}}(\phi)
=+\sqrt{g^2(m^2-\phi^2)/\lambda_\chi}$. More exactly, we choose
$\chi_{i2}=0.01\chi_{\mbox{\scriptsize min}}(0)$ and $\dot{\chi}_{i2}=0$. The
parameters $g$, $m$, and $\lambda_\chi$ should satisfy the
constraints~(\ref{constraint-1}) and (\ref{constraint-2}). We also demand that
$g\lesssim 1$, $\lambda_\chi \lesssim 1$, and $m\lesssim 10^{-2} \,
m_{\mbox{\tiny p}}$. If, for instance, we choose $\lambda_\chi\approx 1$, then
the values of $g$ and $m$ are restricted to the shaded region shown in
Fig.~\ref{fig-constants}. For such values of the parameters the second stage
of evolution lasts typically for about $10^6 \, t_{\mbox{\tiny p}}$ (see
Fig.~\ref{fig-chi-23-etapas}).

During the third stage, the evolution of the Universe is described by the set
of differential equations
\begin{eqnarray}
\frac{\ddot{a}}{a} = -\frac{8\pi}{3m_{\mbox{\tiny p}}^2} \left( \dot{\phi}^2 +
\dot{\chi}^2 - U + \rho_r \right), \label{adotdot3}
\end{eqnarray}
\vspace{-6mm}
\begin{eqnarray}
\ddot{\phi} +3 \frac{\dot{a}}{a} \dot{\phi} +\frac{\partial U}{\partial
\phi}=0,
\end{eqnarray}
\vspace{-6mm}
\begin{eqnarray}
\ddot{\chi} +3 \frac{\dot{a}}{a} \dot{\chi} +\frac{\partial U}{\partial
\chi}=-\Gamma_\chi \dot{\chi},
\end{eqnarray}
\vspace{-6mm}
\begin{eqnarray}
\dot{\rho}_r + 4 \frac{\dot{a}}{a} \rho_r = \Gamma_\chi \dot{\chi}^2,
\label{rhorad3}
\end{eqnarray}
\vspace{-6mm}
\begin{eqnarray}
\left( \frac{\dot{a}}{a} \right)^2 = \frac{8\pi}{3m_{\mbox{\tiny p}}^2} \left(
\frac{\dot{\phi}^2}{2} + \frac{\dot{\chi}^2}{2} + U + \rho_r \right),
\label{friedmann3}
\end{eqnarray}
where $\rho_r$ is the energy density of radiation and $\Gamma_\chi$ is the
dissipative coefficient given by Eq.~({\ref{Gamma_chi}).

Again, we use a Runge-Kutta method to solve the system of differential
equations~(\ref{adotdot3})--(\ref{rhorad3}). Equation~(\ref{friedmann3}) is
used to check the accuracy of the numerical solution. Since any pre-existing
radiation fluid would have been diluted during inflation, we choose the energy
density of radiation at the beginning of the third stage of evolution to be
zero, $\rho_{r,i3}=0$. The dissipation parameter $\mu$ should satisfy the
constraint~(\ref{constraint-3}). For $\lambda_\chi=1$, $m=0.004\,
m_{\mbox{\tiny p}}$, and $g=0.025$ this implies $\mu\gg 10^{-4}$.

As already mentioned above, during the third stage of evolution,
the scalar field $\chi$ oscillates around the minimum of the
potential located at $\chi=0$ (see Fig.~\ref{fig-chi-23-etapas}),
transferring its energy into the radiation fluid (see
Fig.~\ref{fig-rho-3etapa}), thus reheating the Universe. The
reheating temperature depends on the values of the parameters of
the model \cite{bastero-gil-et-al}. For the above-mentioned values
of the parameters, our numerical simulations show that the
reheating temperature is of the order of $10^{14} \mbox{ GeV}$.

\begin{figure}[t]
\includegraphics[width=8.1cm]{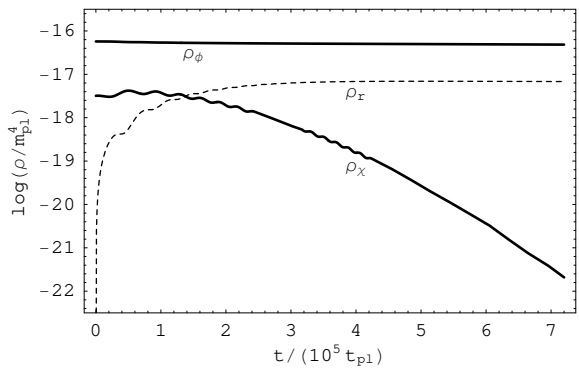}
\caption{Time evolution of the energy densities of radiation,
$\phi$, and $\chi$ at the beginning of the third stage of
evolution (which lasts about $5\times 10^{10} \, t_{\mbox{\tiny
p}}$), for the same values of the parameters as in
Fig.~\ref{fig-w-all-stages}. Energy is transferred from the scalar
field $\chi$ to radiation, while the evolution of the Universe is
dominated by the scalar field $\phi$. Later in this stage, the
energy density of radiation becomes dominant; kination gives place
to a radiation-dominated Universe.} \label{fig-rho-3etapa}
\end{figure}

For low values of the parameter $\mu$, the scalar field $\chi$ oscillates for
a long period of time before its energy is completely transferred to the
radiation fluid. In such cases, because of the difficulty to accurately follow
many oscillations of $\chi$ with our numerical code, a different strategy has
to be used. We start by numerically solving
Eqs.~(\ref{adotdot3})--(\ref{rhorad3}). After a short period of integration,
the terms $g^2 m^2 \chi^2/2$ and $\lambda_\chi\chi^4/4$ in the potential
$U(\phi,\chi)$ become much smaller than $g^2\phi^2\chi^2/2$, implying that the
energy density of the scalar field $\chi$ is given, within a good
approximation, by
\begin{eqnarray}
\rho_\chi \approx \frac{\dot{\chi}^2}{2}+\frac12 g^2 \phi^2 \chi^2.
\end{eqnarray}
Taking into account that $\phi$ remains practically constant during an
oscillation of $\chi$, the oscillations of the latter field can be considered
to be of the simple harmonic type, for which
$\rho_\chi=\langle\dot{\chi}^2\rangle=g^2\phi^2\langle\chi^2\rangle$, where
$\langle\dots\rangle$ denotes the average over one oscillation. Using this
approximation, Eqs.~(\ref{adotdot3})--(\ref{friedmann3}) can be re-written as
\begin{eqnarray}
\frac{\ddot{a}}{a} = -\frac{8\pi}{3m_{\mbox{\tiny p}}^2} \left( \dot{\phi}^2 -
V + \frac12 \rho_\chi + \rho_r \right), \label{adotdot3a}
\end{eqnarray}
\vspace{-6mm}
\begin{eqnarray} \ddot{\phi} +3 \frac{\dot{a}}{a} \dot{\phi}
+\frac{\partial V}{\partial \phi}=-\frac{\rho_\chi}{\phi},
\end{eqnarray}
\vspace{-6mm}
\begin{eqnarray}
\dot{\rho}_\chi + 3 \frac{\dot{a}}{a} \rho_\chi = -\Gamma_\chi \rho_\chi +
\frac{\dot{\phi}}{\phi}\rho_\chi,
\end{eqnarray}
\vspace{-6mm}
\begin{eqnarray}
\dot{\rho}_r + 4 \frac{\dot{a}}{a} \rho_r = \Gamma_\chi \rho_\chi,
\label{rhorad3a}
\end{eqnarray}
\vspace{-6mm}
\begin{eqnarray}
\left( \frac{\dot{a}}{a} \right)^2 = \frac{8\pi}{3m_{\mbox{\tiny p}}^2} \left(
\frac{\dot{\phi}^2}{2} + V + \rho_\chi + \rho_r \right). \label{friedmann3a}
\end{eqnarray}
We then solve these equations, instead of
Eqs.~(\ref{adotdot3})--(\ref{rhorad3}), until the end of the third
stage of evolution.

Let us now turn to the fourth stage of evolution. The scalar field $\chi$ has
already decayed away completely in the previous stage of evolution. Therefore,
Eq.~(\ref{rhorad3}), or Eq.~(\ref{rhorad3a}), can be integrated exactly,
yielding $\rho_r=\rho_{r,0}(a_0/a)^4$, where
$\rho_{r,0}=4.13\times10^{-14}\mbox{ J/m}^3$ and $a_0$ are, respectively,
today's values of the energy density of radiation and of the scale factor. In
this stage of evolution a pressureless matter component is introduced, which
accounts for the usual (baryonic) matter and also for dark matter, with energy
density $\rho_m=\rho_{m,0}(a_0/a)^3$, where
$\rho_{m,0}=2.34\times10^{-10}\mbox{ J/m}^3$ is today's value of the energy
density of matter. Taking all this into account, the equations of motion for
the fourth stage of evolution are given by
\begin{eqnarray}
\hspace{-2mm} \frac{\ddot{a}}{a} = -\frac{8\pi G}{3c^2} \left[ \dot{\phi}^2 -
V + \rho_{r,0} \left( \frac{a_0}{a} \right)^4 + \frac12 \rho_{m,0} \left(
\frac{a_0}{a} \right)^3 \right]\!\! , \label{adotdot4}
\end{eqnarray}
\vspace{-6mm}
\begin{eqnarray}
\ddot{\phi} +3 \frac{\dot{a}}{a} \dot{\phi} +\frac{\partial V}{\partial
\phi}=0,  \label{dotdotphi4}
\end{eqnarray}
\vspace{-6mm}
\begin{eqnarray}
\hspace{-3mm} \left( \frac{\dot{a}}{a} \right)^2 = \frac{8\pi G}{3c^2} \left[
\frac{\dot{\phi}^2}{2} + V + \rho_{r,0} \left( \frac{a_0}{a} \right)^4 +
\rho_{m,0} \left( \frac{a_0}{a} \right)^3 \right]\!\! , \label{friedmann4}
\end{eqnarray}
where $G$ is the gravitational constant and $c$ is the speed of light.

This set of differential equations describes a stage of evolution
in which the Universe is dominated, consecutively, by radiation,
matter and the inflaton/quintessence field $\phi$. As already
mentioned above, in order to obtain agreement with observations
\cite{wmap7}, the parameter $M$ of the potential $V(\phi)$ has to
be chosen such that
\begin{eqnarray}
 \left[ \frac{\dot{\phi}^2}{2} + V(\phi) \right]_{t=t_0}=\rho_{de,0},
\end{eqnarray}
where $\rho_{de,0}=6.20\times10^{-10}\mbox{ J/m}^3$ is today's value of the
energy density of dark energy. Within the present model, the value of $M$ is
typically of the order of $10^{-14}\, m_{\mbox{\tiny p}}$.

Note that the values of $\rho_{de,0}$, $\rho_{m,0}$, and $\rho_{r,0}$ used in
this article imply a value of the Hubble constant of $H_0=71 \mbox{
km}\,\mbox{s}^{-1}\mbox{Mpc}^{-1}$.

\section{Gravitational waves\label{sect-gravitational-waves}}

Gravitational waves are generated in an expanding Universe, giving
rise to a spectrum extending over a wide range of frequencies,
from about $10^{-17}\mbox{ rad/s}$ to about $10^{10}\mbox{ rad/s}$
\cite{grishchuk74, starobinskii79, abbott-harari, Allen88, sahni,
grishchuk-solokhin, allen97}.

In this article, we calculate this spectrum for the hybrid
quintessential inflationary model using the method of the
continuous Bogoliubov coefficients. This method can be summarized
as follows (for details, see
Refs.~\cite{henriques94,henriques-moorhouse-mendes,mendes-henriques-moorhouse,
henriques04,sa-henriques1,sa-henriques2,sa-henriques-potting}).
The number of gravitons at a certain moment of the evolution of
the Universe is given by the squared Bogoliubov coefficient,
$\beta^2=(X-Y)^2/4$, where the functions $X(t)$ and $Y(t)$ are
solutions of the system of differential equations
\begin{eqnarray}
\dot{X} = - i \omega_0  \frac{a_0}{a} Y, \label{XX}
\end{eqnarray}
\vspace{-6mm}
\begin{eqnarray}
\dot{Y} = -\frac{i}{\omega_0} \frac{a}{a_0} \left[ \omega_0^2 \left(
\frac{a_0}{a} \right)^2 - \frac{\ddot{a}}{a}- \left( \frac{\dot{a}}{a}
\right)^2\right]X; \label{YY}
\end{eqnarray}
$a_0$ and $\omega_0$ are today's values of the scale factor and
the gravitational-wave angular frequency, respectively. The above
system of equations is integrated with initial conditions
$X(t_{i1})=Y(t_{i1})=1$, corresponding to the absence of gravitons
at the beginning of the first stage of evolution. The scale factor
$a(t)$ and its first and second derivatives $\dot{a}(t)$ and
$\ddot{a}(t)$ are determined from the evolutionary equations
presented in the previous section. Knowing $\beta^2(t)$, we can
compute the gravitational-wave spectral energy density parameter,
$\Omega_\textsc{gw}$, which is defined as
\begin{eqnarray}
\Omega_\textsc{gw} = \frac{8\hbar G}{3\pi c^5 H^2} \omega^4 \beta^2,
\label{sedp}
\end{eqnarray}
where $H(t)$ is the Hubble parameter. Evaluating all the
quantities in the above expression at the present time, for
angular frequencies ranging from $1.4\times10^{-17}\mbox{ rad/s}$
(corresponding to a wavelength equal, today, to the Hubble
distance) to about $10^{10}\mbox{ rad/s}$ (corresponding to a
wavelength equal to the Hubble distance at the end of the
inflationary period), yields the gravitational-wave energy
spectrum. It is worth emphasizing that this spectrum is fully
determined by the evolution of the scale factor from the beginning
of the inflationary period until the present time. A change in the
behavior of the scale factor, due to a change of the equation of
state of the Universe (see Fig.~\ref{fig-w-all-stages}), leads to
a modification of the slope of the gravitational-wave spectrum at
a certain frequency.

Let us now compute the gravitational-wave spectrum of the hybrid
quintessential inflationary model using the formalism of the continuous
Bogoliubov coefficients. First, we have to specify the values of
$\lambda_\phi$ and $M$, satisfying the observational constraints on inflation
and dark energy, and the values of $g$, $m$, $\lambda_\chi$, and $\mu$,
satisfying the constraints (\ref{constraint-1})--(\ref{constraint-3}). In
Fig.~\ref{fig-espectros-g} we show the full gravitational-wave spectrum for
the hybrid quintessential inflationary model for two different values of the
parameter $g$, corresponding to the minimum and maximum values allowed by the
constraints (\ref{constraint-1}) and (\ref{constraint-2}) for fixed values of
$\lambda_\phi$, $m$, and $\lambda_\chi$.

\begin{figure}[t]
\includegraphics[width=8.1cm]{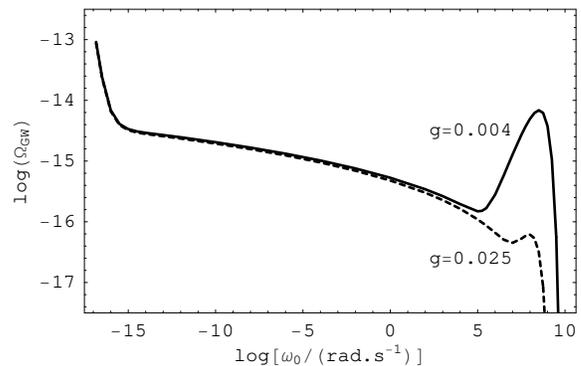}
\caption{Gravitational-wave spectrum for $\lambda_\phi=10^{-13}$, $m=0.004\,
m_{\mbox{\tiny p}}$, $\lambda_\chi=1$, and $\mu=0.1$. The two curves
correspond to the minimum and maximum values of $g$ allowed by the constraints
(\ref{constraint-1}) and (\ref{constraint-2}), namely, $g=0.004$ and
$g=0.025$. In both cases $M$ is of the order of $10^{-14}\, m_{\mbox{\tiny
p}}$.} \label{fig-espectros-g}
\end{figure}

As we can see from Fig.~\ref{fig-espectros-g}, the spectrum is naturally
divided in three regions.

In the low-frequency region, $\omega_0$ ranges from
$1.4\times10^{-17}\mbox{ rad/s}$ to about $2\pi H(t_{m})
a(t_{m})/a_0\approx10^{-15}\mbox{ rad/s}$, where the upper limit
corresponds to today's value of the angular frequency of a
gravitational wave which had a wavelength equal to the Hubble
distance at the time $t_m$ when the energy density of radiation
became equal to the energy density of matter. In this region of
the spectrum, the gravitational-wave spectral energy density
parameter $\Omega_\textsc{gw}$ rises rapidly as $\omega_0$
decreases, due to an extra graviton production during the
transition between the radiation- and the matter-dominated eras
and the subsequent matter- and dark energy-dominated eras. For
$\omega_0 =1.4\times 10^{-17} \mbox{ rad/s}$ the spectrum
satisfies a constraint derived from measurements of the cosmic
microwave background radiation, namely,
$\Omega_\textsc{gw}<1.4\times 10^{-10}$ \cite{allen97}.

In the intermediate-frequency region, the angular frequency ranges
from about $10^{-15}\mbox{ rad/s}$ to $2\pi H(t_{r})
a(t_{r})/a_0$, where $t_{r}$ corresponds to the transition between
the kination and radiation-dominated eras, occurring at the end of
the third stage of evolution. Depending on the duration of the
kination era, the intermediate-frequency region extends up to
values of about $(10^3-10^7) \mbox{ rad/s}$. Note that the
spectrum has, in this region, a nonconstant slope, due to the fact
that inflation is quasi-exponential\footnote{In exponential
inflation (cosmological constant) the slope of the spectrum in the
intermediate-frequency region is zero \cite{starobinskii79}, while
in power-law inflation the slope is constant and negative
\cite{sahni}.}
\cite{sa-henriques-potting,kuroyanagi-chiba-sugiyama}. Several
bounds on the gravitational-wave spectral energy density parameter
should be satisfied in this region of the spectrum, namely, from
timing observations of millisecond pulsars,
$\Omega_\textsc{gw}<4\times10^{-8}$ for $\omega_0=2.5\times
10^{-8}\mbox{ rad/s}$ \cite{jenet-etal}, from Doppler tracking of
the Cassini spacecraft, $\Omega_\textsc{gw}<0.028$ for
$\omega_0=7.5\times 10^{-6}\mbox{ rad/s}$ \cite{armstrong}, and
from the Laser Interferometer Gravitational Wave Observatory
(LIGO), $\Omega_\textsc{gw}<6.9\times10^{-6}$ for
$\omega_0\approx(10^2-10^3)\mbox{ rad/s}$ \cite{LIGO-virgo}.

\begin{figure}[t]
\includegraphics[width=8.1cm]{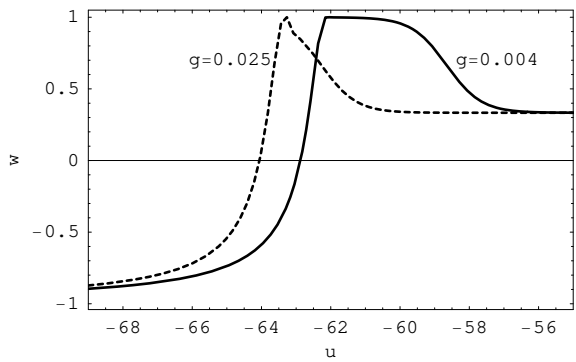}
\caption{Time evolution of the equation-of-state parameter $w=p/\rho$ for
$g=0.004$ and $g=0.025$. The duration of the kination period in the former
case is much longer, implying a higher peak in the gravitational-wave
spectrum. In this figure the values of the parameters are the same as in
Fig.~\ref{fig-espectros-g} and $u=-\ln(1+z)$, where $z$ is the redshift.}
\label{fig-w-parcial-g}
\end{figure}

In the high-frequency region of the spectrum, ranging from
$(10^3-10^7) \mbox{ rad/s}$ to about $10^{10} \mbox{ rad/s}$, one
observes a sharp rise of the gravitational-wave spectral energy
density parameter $\Omega_\textsc{gw}$. This rather interesting
feature of the spectrum is due to the existence of a kination
period (with equation of state $p=w\rho$, $w=1$) between the end
of inflation and the beginning of the radiation-dominated era,
whose duration determines the height of the peak
\cite{giovannini98}.

Our numerical simulations show that the duration of the kination period
increases as the value of the parameter $g$ decreases (see
Fig.~\ref{fig-w-parcial-g}). This can be easily understood as follows. If one
decreases $g$ (for fixed values of $m$ and $\lambda_\chi$), the value of
$\chi$ at the minimum of the potential also decreases (recall that
$\chi_{\mbox{\scriptsize min}} = \pm \sqrt{g^2(m^2-\phi^2)/\lambda_\chi}\, $),
meaning that less energy is acquired by this field as it oscillates around the
minimum during the second stage of evolution. As a consequence, less energy is
available to be transferred, during the third stage of evolution, from the
scalar field $\chi$ to the radiation fluid, leading to a lower reheating
temperature. This, in turn, implies that after the complete decay of $\chi$ a
longer period of time is required for the energy density of radiation to
become greater than the kinetic energy of the scalar field $\phi$, i.e., the
beginning of the radiation-dominated era is delayed and, consequently, the
kination period becomes longer. In short, we could say that reheating becomes
less efficient as $g$ decreases.

As we have seen in Sect.~\ref{sect-hybrid-quint-inflation}, $g$ is not a free
parameter, its value is bounded from above by the condition that the motion of
$\phi$ is not affected by $\chi$ [see Eq.~(\ref{constraint-2})] and from below
by the condition that the scalar field $\chi$ responds quickly enough to
changes in the potential $U(\phi,\chi)$, rolling down toward one of the new
minima and oscillating around it [see Eq.~(\ref{constraint-1})]. Therefore,
within the hybrid quintessential inflationary model, the duration of the
kination period and, consequently, the height of the peak in the
gravitational-wave spectrum, cannot be freely adjusted, they are limited by
the allowed values of the parameter $g$.

What in the previous two paragraphs was said relatively to the
dependence of the gravitational-wave spectrum on the value of $g$
could also be said, with the necessary adaptations, about the
parameters $m$ and $\lambda_\chi$. The dependence of the spectrum
on the parameter $m$ is illustrated in Fig.~\ref{fig-espectros-m}.
There, two spectra are shown, corresponding to the minimum and
maximum values of $m$ allowed by the constraints
(\ref{constraint-1}) and (\ref{constraint-2}) for fixed values of
$\lambda_\phi$, $g$, and $\lambda_\chi$. For both values of $m$,
the parameter $\mu$ is chosen such that constraint
(\ref{constraint-3}) is satisfied. As we see in this figure, the
height of the peak, located at high frequencies, increases as the
parameter $m$ decreases. In Fig.~\ref{fig-espectros-lambdachi},
three gravitational-wave spectra are shown for different values of
$\lambda_\chi$ and fixed values of $\lambda_\phi$, $g$, $m$ and
$\mu$. As expected, the height of the peak increases as the
parameter $\lambda_\chi$ increases. In all cases considered above,
the parameter $M$ is of the order of $10^{-14}\, m_{\mbox{\tiny
p}}$.

\begin{figure}[t]
\includegraphics[width=8.1cm]{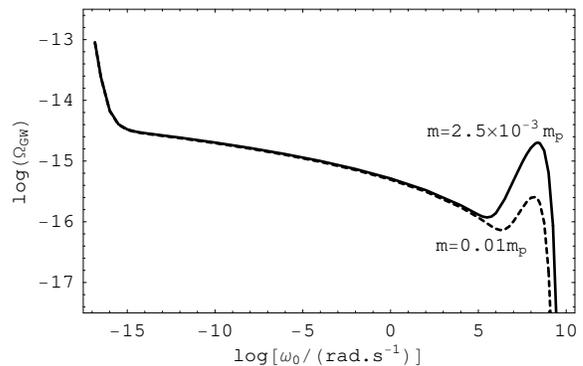}
\caption{Gravitational-wave spectrum for $\lambda_\phi=10^{-13}$, $g=0.01$,
$\lambda_\chi=1$, and $\mu=1$. The two curves correspond to the minimum and
maximum values of $m$ allowed by the constraints (\ref{constraint-1}) and
(\ref{constraint-2}), namely, $m=2.5\times10^{-3}\, m_{\mbox{\tiny p}}$ and
$m=0.01\, m_{\mbox{\tiny p}}$. In both cases $M$ is of the order of
$10^{-14}\, m_{\mbox{\tiny p}}$.} \label{fig-espectros-m}
\end{figure}

\begin{figure}[t]
\includegraphics[width=8.1cm]{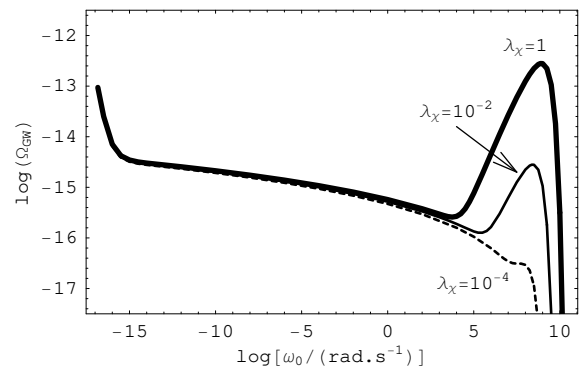}
\caption{Gravitational-wave spectrum for $\lambda_\phi=10^{-13}$,
$g=6.3\times10^{-4}$, $m=0.01\, m_{\mbox{\tiny p}}$, and $\mu=0.1$. The three
curves correspond to $\lambda\chi=10^{-4},\, 10^{-2},\, 1$. The minimum value
of $\lambda_\chi$, allowed by constraints (\ref{constraint-2}) and
(\ref{constraint-3}), is $1.5\times10^{-5}$. In all cases $M$ is of the order
of $10^{-14}\, m_{\mbox{\tiny p}}$.} \label{fig-espectros-lambdachi}
\end{figure}

Let us now turn to the analysis of the influence of the dissipation parameter
$\mu$ on the duration of the kination period and, consequently, on the height
of the high-frequency peak of the gravitational-wave spectrum. Remember that
the parameter $\mu$ is bounded from below by the condition that the motion of
$\phi$ is not affected by $\chi$ [see Eq.~(\ref{constraint-3})]. We also
require that this parameter is smaller than a critical value,
$\mu_{\mbox{\scriptsize crit}}$, above which the oscillatory motion of the
scalar field $\chi$ during the third stage of evolution would become
over-damped. For values of the dissipation parameter $\mu$ lying in the
interval $0.2 \, \lambda_\phi^{1/6} \left( m_{\mbox{\tiny p}}/m \right)^{2/3}
g^{5/3}\lambda_\chi^{-1} \ll \mu < \mu_{\mbox{\scriptsize crit}}$, the
duration of the kination period decreases as $\mu$ decreases. This is due to
the fact that for smaller values of $\mu$ the energy transfer of the scalar
field $\chi$ to the radiation fluid proceeds slower, with the consequence that
the energy density of radiation begins to decrease as $a^{-4}$ later.
Therefore, after the complete decay of the field $\chi$, the energy density of
radiation is higher in the case of small $\mu$ and the time it takes for this
energy density to dominate the dynamics of the Universe is shorter, implying
that the kination period is also shorter. In Fig.~\ref{fig-espectros-mu} we
plot three gravitational-wave spectra for different values of $\mu$, holding
fixed the other parameters of the model. We see that the height of the peak in
the high-frequency region of the spectrum increases as the dissipation
parameter increases, achieving its maximum value for
$\mu=\mu_{\mbox{\scriptsize crit}}$.

\begin{figure}[t]
\includegraphics[width=8.1cm]{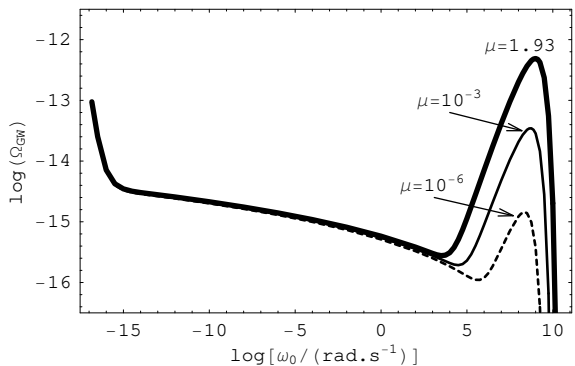}
\caption{Gravitational-wave spectrum for $\lambda_\phi=10^{-13}$,
$g=6.3\times10^{-4}$, $m=0.01\, m_{\mbox{\tiny p}}$, and $\lambda\chi=1$. The
three curves correspond to $\mu=\mu_{\mbox{\scriptsize crit}}=1.93$, $\mu=
10^{-3}$, and $\mu= 10^{-6}$. Constraint (\ref{constraint-3}) requires that
$\mu\gg 10^{-7}$. In all cases $M$ is of the order of $10^{-14}\,
m_{\mbox{\tiny p}}$.} \label{fig-espectros-mu}
\end{figure}

To complete our analysis of the influence of the various
parameters of the potential $U(\phi,\chi)$ on the
gravitational-wave spectrum, we point out that the parameter
$\lambda_\phi$ determines the overall vertical displacement of the
spectrum (see Fig.~\ref{fig-espectros-phi}). Note that the value
of $\lambda_\phi$ also affects the height of the peak, contrarily
to the situation in original quintessential inflationary model,
where only the low- and intermediate-frequency regions of the
spectrum were pushed down (up) by an decrease (increase) of
$\lambda_\phi$ (see Fig.~4 of Ref.~\cite{giovannini99}).

\begin{figure}[t]
\includegraphics[width=8.1cm]{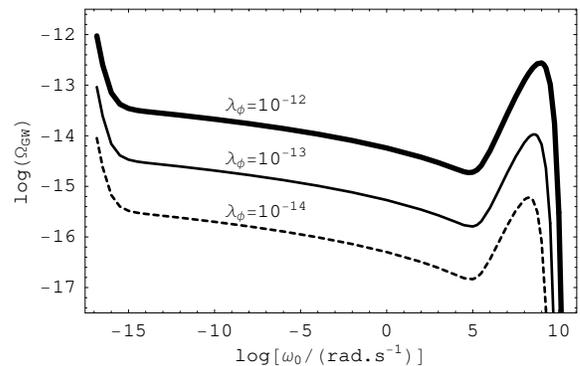}
\caption{Gravitational-wave spectrum for $g=0.002$, $m=0.01\, m_{\mbox{\tiny
p}}$, $\lambda\chi=1$, and $\mu= 0.1$. The three curves correspond to
$\lambda_\phi=10^{-14},\,10^{-13},\, 10^{-12}$. In all cases $M$ is of the
order of $10^{-14}\, m_{\mbox{\tiny p}}$.} \label{fig-espectros-phi}
\end{figure}

To finish this section, let us point out that, for the allowed
values of the different parameters of the model, all
gravitational-wave spectra satisfy, by far, the integral bound
imposed by primordial nucleosynthesis \cite{allen97, maggiore},
\begin{eqnarray}
 \int\limits_{\omega_n}^{\infty} \Omega_\textsc{gw} d\ln\omega <
 1.1\times10^{-5}, \label{primordial-nucleosynthesis}
 \end{eqnarray}
where $\omega_n\approx 10^{-9}\mbox{ rad/s}$.

\section{Conclusions\label{sect-conclusions}}

In this work we have investigated the generation of gravitational waves in the
hybrid quintessential inflationary model. In this model, recently proposed by
Bastero-Gil \emph{et al.}~\cite{bastero-gil-et-al}, a unified description of
inflation and dark energy is achieved with a single scalar field $\phi$
playing both roles of inflaton and quintessence, while reheating takes place
in the usual way, through the complete decay of another scalar field, $\chi$,
rather than by gravitational particle production, as in the original
quintessential inflationary model of Peebles and Vilenkin
\cite{peebles-vilenkin}.

The hybrid-like quintessential inflationary potential,
$U(\phi,\chi)$, contains several parameters which are constrained
by measurements of the cosmic microwave background radiation and
dark energy, as well as by certain restrictions imposed on the
behavior of the scalar field $\chi$, namely, that it responds
quickly enough to changes in the potential $U(\phi,\chi)$ and that
it does not influence significantly the evolution of the
inflaton/quintessence field~$\phi$.

Another relevant parameter of the model is the dissipation parameter which
regulates the rate at which energy is transferred from the scalar field $\chi$
to radiation. This parameter is constrained from below by the condition that
the motion of $\phi$ is not influenced significantly by $\chi$ and from above
by the requirement that the oscillations of $\chi$ around the minimum of the
potential are not over-damped.

For the allowed values of the above-mentioned parameters, we have
calculated the full gravitational-wave energy spectrum using the
method of continuous Bogoliubov coefficients. Such spectra carry,
at high frequencies, a clear signature of quintessential
inflation, namely, a sharp rise of the gravitational-wave spectral
energy density parameter $\Omega_\textsc{gw}$. This distinctive
feature of the spectrum is due to the existence of a
post-inflationary kination period, during which the kinetic energy
of the inflaton field dominates the evolution of the Universe
\cite{giovannini98}. As we have shown in this paper, the duration
of the kination period depends crucially on the parameters of the
hybrid-like potential, $g$, $m$ and $\lambda_\chi$, as well as on
the dissipation parameter $\mu$. Namely, if $g$ and $m$ decrease
($\lambda_\chi$ increases), reheating becomes less efficient, with
the consequence that the kination period becomes longer. In this
case, the peak in the high-frequency region of the
gravitational-wave spectrum becomes more pronounced (see
Figs.~\ref{fig-espectros-g}, \ref{fig-espectros-m} and
\ref{fig-espectros-lambdachi}). Similarly, if $\mu$ increases, the
duration of the kination period also increases, implying a higher
peak in the spectrum (see Fig.~\ref{fig-espectros-mu}).

In the original quintessential inflationary model \cite{peebles-vilenkin}, the
kination period is quite long, implying a pronounced peak in the
high-frequency region of the gravitational-wave spectrum \cite{giovannini99}.
In order to avoid a conflict with the integral bound imposed by primordial
nucleosynthesis [see Eq.~(\ref{primordial-nucleosynthesis})], in the original
model one has to require that the number of scalar fields reheating the
Universe is greater than about $10^2$ \cite{peebles-vilenkin}, meaning that
the minimal Grand Unification Theory is not enough to accommodate
quintessential inflation. This situation contrasts with the one occurring in
the hybrid quintessential inflationary model. As we have shown in this paper,
for the allowed values of the parameters of the model, the
primordial-nucleosynthesis bound is satisfied, by far, by any
gravitational-wave spectrum. Indeed, even when all parameters of the model are
pushed to their extreme values, the gravitational-wave spectral energy density
parameter $\Omega_\textsc{gw}$ does not grow above $10^{-12}$. This important
difference between the gravitational-wave spectrum of the original model and
its hybrid variant is due to the fact that in the latter the duration of the
kination period is much shorter, implying a smaller peak in the
gravitational-wave spectrum.

The values of the different parameters also determine the frequency at which
the gravitational-wave spectral energy density parameter $\Omega_\textsc{gw}$
starts to rise at the high-frequency region of the spectrum. For extreme
values of the parameters, this frequency can be as low as $10^2\mbox{ Hz}$.
Despite the fact that this frequency falls within the frequency range at which
terrestrial laser-interferometer gravitational-wave detectors operate, a
detection in a near future is improbable, due to the very low value of
$\Omega_\textsc{gw}$, about $10^{-15}-10^{-16}$.

In the gravitational-wave energy spectrum, the maximum of the peak
is firmly located in the MHz-GHz region and, within the hybrid
quintessential inflationary model, can reach values of about
$\Omega_\textsc{gw} \simeq 10^{-12}$. A search for the primordial
gravitational-wave background at this range of frequencies was
initiated a few years ago, with both microwave-cavity detectors
\cite{bernard-etal,cruise-ingley} and interferometric detectors
\cite{akutsu-etal}. We deem the development of such detectors to
be of the utmost importance, since they will open a direct window
to the very early Universe, allowing us to test our theoretical
ideas about the inflationary and post-inflationary epochs.

\vspace{2mm} {\color{white}.}

\begin{acknowledgments}
The authors thank R.~Potting for interesting discussions. This
work was supported in part by the Funda\c{c}\~ao para a Ci\^encia
e a Tecnologia, Portugal.
\end{acknowledgments}

\end{document}